\documentclass[conference]{IEEEtran}

\usepackage{cite}
\usepackage{amsmath,amssymb,amsfonts,amsthm}
\usepackage{algorithmic}
\usepackage{algorithm}
\usepackage{graphicx}
\usepackage{textcomp}
\usepackage{xcolor}
\usepackage{url}
\usepackage{verbatim}
\usepackage{tabularray}
\usepackage{csquotes}

\usepackage{balance}

\usepackage{color, colortbl}
\definecolor{LightGray}{gray}{0.9}

\makeatletter
\newcommand{\linebreakand}{%
  \end{@IEEEauthorhalign}
  \hfill\mbox{}\par
  \mbox{}\hfill\begin{@IEEEauthorhalign}
}
\makeatother

\theoremstyle{definition}
\newtheorem{definition}{Definition}

\begin{document}

\title{Lightning Fast Matching Dependency Discovery with Desbordante}
\date{}

\author{
\IEEEauthorblockN{Alexey Shlyonskikh,
Michael Sinelnikov,
Daniil Nikolaev,
Yurii Litvinov,
George Chernishev
}
\IEEEauthorblockA{Saint-Petersburg University \\ Saint-Petersburg, Russia \\ {\{shlyonskikh.alexey, michael.a.sinelnikov, daniil.vl.nikolaev\}}@gmail.com, {\{y.litvinov, g.chernyshev\}}@spbu.ru\\}}

\maketitle

\begin{abstract}

Matching dependency is a generalization of the functional dependency concept, which allows users to apply custom similarity functions for matching individual attributes. Matching dependencies have a wide range of applications for solving various data quality problems, such as entity resolution, data deduplication, data integration, schema matching, and many more. However, their discovery is a very computationally intensive problem, which limits their practical application.

In this paper, we describe a number of optimization techniques for HyMD~--- currently the state-of-the-art algorithm for the discovery of matching dependencies. These optimizations belong to both technical and scientific domains. The most important of them are: 1) a new sampling technique, 2) a faster generalization lookup technique, and 3) an improved representation of a dependency. The first one aims to raise the efficiency of inference from record pairs, while the last two are designed to speed up lattice-related operations.

To evaluate our optimizations, we implemented our version of HyMD in Desbordante, an open-source high-performance data profiler. Experiments demonstrated that they allow for a speedup of more than 40x over the state-of-the-art implementation on average, reaching a speedup greater than 170x in some cases.

Finally, the improved version of HyMD is ready to use by anyone. It comes with bidirectional Python integration, which allows calling the C++ algorithm implementation from Python programs while allowing users to supply their custom matching functions. 
\end{abstract}

\section{Introduction}

Data profiling~\cite{10.5555/3312004} involves extracting various types of information from data. Such information can range from simple statistics (e.g. the number of columns in a table) to complex facts, indicating the presence of various patterns in data. We refer to the first type of extraction~\cite{DBLP:journals/corr/abs-2301-05965} as \textbf{naive profiling} and to the second one as \textbf{science-intensive profiling}.

Science-intensive data profiling uses computationally costly algorithms to extract sophisticated patterns from the data, such as functional dependencies~\cite{DBLP:journals/VLDB/PapenbrockEMNRSZN15}, inclusion dependencies~\cite{10.1145/3357384.3357916}), association rules~\cite{10.5555/2677098}, algebraic constraints~\cite{10.5555/1315451.1315509}, inferred semantic data types~\cite{10.1145/3292500.3330993}, and others. These patterns represent knowledge about data and can be used to perform various tasks in many domains: data quality, data integration, database management, query optimization, database reverse engineering, and many more.

One of the well-known patterns is matching dependency (MD), which is defined on tables. Informally, it states that for all record pairs where some attribute values are \enquote{similar}, some other attributes are also \enquote{similar}.

\begin{table}
\centering
\caption{Airline flights}
\label{tbl:routes}
\begin{tabular}{l|l|l|l|c}
id & Source & From             & To               & Distance (km)  \\ 
\hline \hline
1        & ac1     & Saint-Petersburg & Helsinki         & 315     \\
2        & ac2     & St-Petersburg    & Helsinki         & 301     \\
3        & ac2     & Moscow           & St-Petersburg    & 650     \\
4        & ac2     & Moscow           & St-Petersburg    & 638     \\
5        & ac1     & Moscow           & Saint-Petersburg & 670     \\
6        & ac1     & Moscow           & Yekaterinburg    & 1417         
\end{tabular}
\end{table}

Consider an example presented in Table~\ref{tbl:routes}. It lists airline routes, which ended up in a single table after data from two airline carriers (\textit{ac1} and \textit{ac2}) was merged. This data contains the following MD:

$$From_{0.75} To_{0.75} \rightarrow Distance_{0.9}.$$

The columns involved have different data types~--- string and integer. Therefore different metrics should be used to specify similarity. For $From$ and $To$ we employ Jaccard distance over 1-grams, and for $Distance$ we use euclidean distance.

Basically, it says the following: for each pair of flights where $From$ and $To$ attributes are close to each other, the difference between their $Distance$ will be less than 10\%. We set the threshold to consider them close to $0.75$. This works well for the example table, as flights 1 and 2 are close to each other since their $To$s are similar to each other (equal in this case), while $From$ are within the specified threshold as $dist_{Jacc}(\text{Saint-Petersburg}, \text{St-Petersburg}) = 10 / 13 \approx 0.76 > 0.75$. The same holds true for flights 3--5, considered in a pairwise manner. Flight number 6 does not have flight records with similar attribute values, so it is not checked. With this dependency at hand, a user can clean and deduplicate the respective data.

Matching dependencies are one of the most important patterns as they possess significant expressive power, comparable only to differential dependencies~\cite{song2011differential}. They have sound theoretical base, such as formal semantics and inference rules introduced in~\cite{fan2009reasoning}, theory for data cleaning and consistent query answering problems studied in detail in~\cite{bertossi2011data}. In practice matching dependencies can be used for solving many data quality problems: record matching, entity resolution, data deduplication, data integration, data schema matching, defining and maintaining integrity constraints, and many more. 

Recently, the MDedup~\cite{mdedup} system highlighted the usefulness of MDs for data deduplication. It is a fully automatic system for detection of duplicates which requires no domain knowledge or training data.

The central issue of matching dependency discovery is that these algorithms have exponential time complexity. Therefore, sophisticated optimizations are required to ensure practical viability. At the same time, existing implementations are merely prototypes intended to demonstrate the viability of the approach. Moreover, authors consider only the algorithmic side of the problem, while ignoring the implementation side completely. In the aforementioned circumstances, it is essential to make use of any opportunities to speed up dependency discovery. Finally, existing implementations fail to cater to user needs: they offer few or no configuration options, are hard to set up and run, and do not provide Python integration, as they are stand-alone programs.

We aim to address these issues by implementing our version of the HyMD algorithm as a part of the Desbordante platform. Desbordante (Spanish for \textit{boundless})~\cite{DBLP:journals/corr/abs-2301-05965, DBLP:journals/corr/abs-2307-14935} is a science-intensive open-source data profiling tool with a focus on implementation performance. The tool is also able~\cite{DBLP:conf/fruct/SmirnovCSBC23, DBLP:conf/fruct/KuzinSPSFC24} to discover other types of dependencies, including functional dependencies (exact and approximate), conditional functional dependencies, order dependencies, and many others. The full list can be found in the repository (https://github.com/desbordante/desbordante-core/). Desbordante is not only capable of dependency discovery, but it can also validate data dependencies, while providing users with additional information useful in their applications. Finally, Desbordante is easy to set up and run, it offers a rich set of configuration options, and all information related to discovered or validated patterns is available inside Python programs via Desbordante Python bindings.

Overall, the contribution of this paper is the following:
\begin{itemize}
    \item A comprehensive study of HyMD~--- the state-of-the-art algorithm for the discovery of matching dependencies.
    \item The first industrial-grade implementation for discovery of matching dependencies. It is fully open-source and comes with a Python interface.
    \item A number of optimization techniques which allows to achieve up to a 170x speedup (greater than 40x on average) compared to existing prototypes.
    \item A discussion of additional optimizations that can improve performance further.
\end{itemize}

This paper is organized as follows. In Section~\ref{sec:background}, we introduce the necessary definitions. Next, in Section~\ref{sec:relwork}, we present related work concerning matching dependencies. In Section~\ref{sec:algorithm}, we describe the existing algorithm, and in Section~\ref{sec:improvements}, we present our improvements. Their evaluation is discussed in Section~\ref{sec:eval}, with further possible improvements discussed in Section~\ref{sec:future}, and Section~\ref{sec:concl} concluding the paper.

\section{Background}\label{sec:background}

Let's start by introducing the core definitions for MDs. For the sake of clarity, we will adhere to the notation used in~\cite{hymd}.

\begin{definition}
    A \textbf{similarity measure} $\approx$ is a function defined for pairs of values, with a result belonging to the $[0.0, 1.0]$ range. Semantically, $0.0$ indicates that the values are totally dissimilar, while $1.0$ indicates maximum similarity.
\end{definition}

Examples of such measures are Levenshtein distance normalized by the maximum string length, Jaccard distance, and many others.

\begin{definition}
    A \textbf{decision boundary} $\lambda$ (or $\rho$) is a possible value of a similarity measure and is used to determine whether two values are similar or dissimilar. If the result of a similarity measure is greater than or equal to the decision boundary, the values are considered to be similar.
\end{definition}

To define a matching dependency, we consider two (potentially identical) relations $R$ and $S$, attributes $A_i \in R$, $B_i \in S$, and a set of \textbf{column matches} $C_i = (A_i, B_i, \approx_i) \in R \times S \times \approx$. 
Column match is a formalization of notion \enquote{values in columns $A_i$ and $B_i$ can be similar to each other relative to metric $\approx_i$}. 
Note that when $R = S$, column matches typically map columns onto themselves in practical use, but when $R \neq S$, a more complex matching policy can be used. It is also worth noting that there are no restrictions on the similarity measure, therefore the columns $A_i$ and $B_i$ in column match $C_i$ can belong to different domains.

Now, if we add decision boundaries to column matches, we can formally define matching dependencies:

\begin{definition}
    A \textbf{matching dependency} $\varphi(\lambda, \rho_j)$ over a set of column matches $C \subseteq R \times S \times \approx$, decision boundaries for the left-hand side $\lambda = \{\lambda_1, \ldots, \lambda_m\}$ (where $m=|C|$) and a decision boundary for the right-hand side $\rho_j$ is 

    $$(\bigwedge\limits_{i=1}^mR[A_i] \approx_{i, \lambda_i} S[B_i]) \rightarrow R[A_j] \approx_{j, \rho_j} S[B_j]$$
\end{definition}

Note that originally~\cite{fan2009reasoning} matching dependencies were defined with a list of attributes on the right-hand side. However, later it was shown~\cite{hymd} that such form is equivalent to a set of matching dependencies with a single attribute on the right-hand side. We will therefore use the simpler notation with a single attribute following~\cite{hymd}.

\begin{definition}
    Given instances $r$ and $s$ of relations $R$ and $S$, we say that a pair of records $(r_k, s_l) \in r \times s$ \textbf{matches the LHS of MD $\varphi$} \emph{iff} 
    $$\bigwedge\limits_{i=1}^m r_k[A_i] \approx_i s_l[B_i] \geq \lambda_i.$$

    Similarly, we say that a pair of records $(r_k, s_l)$ \textbf{matches the RHS of MD $\varphi$} \emph{iff}   
    $$r_k[A_j] \approx_{j} s_l[B_j] \geq \rho_j$$
\end{definition}

\begin{definition}
    Matching dependency $\varphi$ \textbf{holds} for instances r and s of relations R and S \emph{iff} $\forall (r_k, s_l) \in (r \times s)$ if LHS $\varphi$ matches $(r_k, s_l)$ then RHS $\varphi$ matches $(r_k, s_l)$, or $\forall (r_k, s_l) \in (r \times s)$ the following holds:
    \begin{equation*}    
    \bigwedge\limits_{i=1}^mr_k[A_i] \approx_{i, \lambda_i} s_l[B_i] \implies r_k[A_j] \approx_{j, \rho_j} s_l[B_j]
    \end{equation*}
    
\end{definition}

With this definition, there are infinitely many MDs for all non-empty sets of column matches. To meaningfully search for MDs on datasets, we need to narrow down the search space.

\begin{definition}
    Given instances $r$ and $s$ of relations $R$ and $S$ and a column match $C_i = (A_i, B_i, \approx_i) \in R \times S \times \approx$, a decision boundary $\lambda_i$ is called \textbf{natural} \emph{iff} 

    $$\exists (r_k, s_l) \in r \times s : r_k[A_i] \approx_i s_l[B_i] = \lambda_i.$$
\end{definition}

Thus, a decision boundary is natural when it is an actual value of similarity between some values in the analyzed dataset. With this, the search space becomes finite, but still very large.

\begin{definition}
    We say that $\varphi(\lambda, \rho_j)$ \textbf{subsumes} $\varphi'(\lambda', \rho'_j)$ \emph{iff} 
    $$\forall i \in [1, m]\; \lambda_i \leq \lambda'_i \wedge \rho_j \geq \rho'_j.$$

    We will denote this fact as $\varphi(\lambda, \rho_j) \preceq \varphi'(\lambda', \rho'_j)$.

    We also say that $\varphi$ \textbf{generalizes} $\varphi'$ and $\varphi'$ \textbf{specializes} $\varphi$.
\end{definition}

If $\varphi(\lambda, \rho_j) \preceq \varphi'(\lambda', \rho'_j)$ and $\varphi(\lambda, \rho_j)$ holds, then $\varphi'(\lambda', \rho'_j)$ holds as well. Therefore, it is only necessary to discover MDs that are not subsumed by any other, as others can be inferred from them.

\begin{definition}
    MD $\varphi(\lambda, \rho_j)$ is \textbf{minimal} \emph{iff} 
    $$\nexists \varphi': \varphi' \preceq \varphi\ \wedge\ \varphi' \neq \varphi$$
\end{definition}

Outputting only minimal MDs reduces result set size and discovery time significantly. Still, not all MDs are needed in practice.

\begin{definition}
    MD $\varphi(\lambda, \rho_j)$ is called \textbf{trivial} \emph{iff} $\lambda_j \geq \rho_j$.
\end{definition}

Trivial dependencies always hold, independent of data, thus we do not need to discover them.

However, not all non-trivial MDs are actually useful, since there are possibly thousands of MDs even in relatively small datasets, and many of them are not very informative (e.g. have low similarity thresholds) or purely coincidental and are discovered only because there is no counterexample present in a dataset. We will therefore briefly list interestingness criteria from~\cite{hymd} that allow us to filter out such MDs:

\begin{itemize}
    \item \textbf{Cardinality} of MD $\varphi(\lambda, \rho_j)$ is the number of non-zero decision boundaries in the LHS. MDs with low cardinality are \enquote{more valuable}, since they are easier to interpret and are more likely to actually be present in the domain the underlying data belongs to.
    
    \item \textbf{Support} is the number of record pairs in the datasets that match the LHS of an MD. MDs with high support are \enquote{more valuable}, since high support means that MD holds for more records in a dataset. Note that if an MD is defined over a single table ($r = s$) then its support is, in the typical case, at least $|r|$ , since each record is usually matched to itself, but if $r \neq s$, then support may equal zero. Also, note that an MD with zero support can be valid, but uninformative, as it does not provide any information regarding the actual dependencies in the data.

    \item \textbf{Disjointness} is a property of an MD which is true \emph{iff} $\lambda_j = 0$ when $j$ is the attribute from an RHS. Authors of~\cite{hymd} have found no practical use for non-disjoint MDs, so they suggest to prune them from the search space entirely.

    \item \textbf{Decision boundaries} with very low thresholds are uninteresting, because they are likely to be ``accidental'' and, at the same time, are not informative. Natural decision boundaries that are very close to one another can be \enquote{merged}, as the corresponding MDs will likely provide the data analyst with the same insights about the analyzed data. The number of decision boundaries dramatically affects the search space size, and the resulting performance of the algorithm. Authors of~\cite{hymd} suggest using 0.7 as the minimal decision boundary value as well as limiting the number of considered boundaries. 
    
\end{itemize}

\section{Related Work}\label{sec:relwork}

Matching dependencies were first introduced in~\cite{fan2008dependencies} and properly formalized in its expanded version~\cite{fan2009reasoning}. However, both of these works did not present an algorithm for MD discovery. In those works, matching dependencies were defined over two potentially different tables and used the notion of \emph{matching operator}, which is as follows: if an LHS of some matching dependency holds true for a given pair of tuples, then RHS attributes are considered to \emph{match}. This match intuitively means that they denote the same domain object. These works also introduce a notion of relative candidate key (RCK)~--- a special form of MD that is used to uniquely identify a tuple in a table. It is similar to the traditional key in relational databases, but is able to tolerate imprecisions and errors in data.

The main focus of early works on MDs was on record matching and data cleaning, with the assumption that MDs were already known, having been supplied by a domain expert.

The first discovery algorithm for matching dependencies was proposed by Song and Chen~\cite{song2009discovering} (note that it was a conference paper, and that there exists an extended journal version~\cite{song2013efficient}; therefore, we refer all readers interested in the Song and Chen algorithm to this paper). This algorithm solves a much simpler task than the later ones, e.g. HyMD~\cite{hymd} and its modification that is discussed in our work. Their problem formulation was as follows: given an already known set of LHS attributes and an RHS attribute of a matching dependency, find decision boundaries for each involved attribute. Note that they consider only MDs defined over a single table.

The original algorithm differs from HyMD in that it searches only for approximate decision thresholds, whereas the latter identify exact thresholds. The Song and Chen algorithm is based on a similarity statistics table that contains the probability of two random pairs of tuples from the relation having a specified similarity threshold vector. For this similarity statistics table to be non-trivial and feasible for in-memory storage, they discretize similarity values (for example, mapping them to natural numbers  from 0 to 10, where 0 is equal and 10 is completely different). Then authors iterate over the similarity statistics table, pruning MDs by support and \emph{confidence} (a fraction of tuple pairs on which an MD holds). Additionally, during this iteration, the algorithm performs some basic lattice traversal pruning. Experiments demonstrate that, even with all of the optimizations, it is still not feasible to use this algorithm for real-world datasets, as it requires seconds to process datasets with only several thousands of records and a dozen of attributes despite solving a much simpler task than modern algorithms.
However, this algorithm can discover approximate MDs (i.e. MDs that only hold on a specified fraction of tuples), which makes it useful for inconsistency detection purposes.

To the best of our knowledge, the latest and fastest algorithm for MD discovery is HyMD~\cite{hymd}. It is an adaptation of HyFD~\cite{hyfd}, an algorithm for exact functional dependencies discovery from the same authors. Both algorithms are named \enquote{hybrid} because they combine lattice search with inference of dependencies from pairs of records. Lattice search builds candidate dependencies and validates them against the data in the table, discarding candidates that will never hold, and pruning those that are non-minimal or fail to satisfy interestingness criteria. Lattice search is exponential in the number of table attributes, but performs very well on \enquote{long} datasets.
Inference from record pairs enumerates pairs of tuples in a table and invalidates dependencies that do not hold on the current pair. This approach is quadratic in the number of tuples of a dataset. In hybrid algorithms lattice search and inference from record pairs work together and support each other~--- lattice search provides \enquote{interesting} pairs for inference, and inference invalidates dependencies, thus removing entire branches from lattice search. HyFD and HyMD switch lattice search and inference phases when the current phase becomes ineffective.

HyMD, as presented in~\cite{hymd}, already uses advanced optimization techniques: algorithmic, such as pruning search space by \enquote{interestingness} of MDs, and purely technical, such as employing dictionary compressed records and similarity indexes. 
Our work differs from this as it presents an industrial-grade implementation of the HyMD algorithm, written in C++ and containing even more technical optimizations, while also being ready for integration through the provided Python bindings.

There are some recent advancements on fast discovery of other kinds of related dependencies, such as differential dependencies~\cite{kuang2024efficient}. Differential dependencies (DDs) are similar to MDs in their ability to capture similarity between values, but they are more expressive as they allow users to require values to be dissimilar. For example, the fact \enquote{if arguments of cryptographic hash function are close, results shall significantly differ} can be expressed as a DD with metric functions that capture \enquote{close} and \enquote{differ} semantics. MDs specified over a single relation can be considered a special case of DDs, but experimental results reported in~\cite{kuang2024efficient} show that MDs can be discovered more efficiently, which is not surprising as DDs are more expressive. On the other hand, the possibility of adapting a hybrid algorithm for DD discovery can be worthy of further study.

An overview of other types of data dependencies and their relations can be found in~\cite{song2020data}.

\section{Algorithm Description}\label{sec:algorithm}

\subsection{Preliminaries}

We will now provide a brief overview of building blocks of the HyMD algorithm, while details and in-depth explanations can be found in~\cite{hymd}.

There are two general ways of discovering data dependencies relevant to HyMD: \emph{lattice traversal} and \emph{inference from record pairs}.

Both rely on a defined product order on the dependencies to limit the number of dependencies considered. At any point in the algorithms, only minimal dependencies according to this order that are assumed to hold are stored. The partially ordered set of dependencies is a bounded lattice and the structure that stores the dependencies assumed to be holding is called a lattice.

Lattice traversal algorithms walk through the space of possibly valid dependencies in accordance with the aforementioned order, checking whether each candidate holds on a dataset. A prominent example of this class of algorithms is the TANE~\cite{tane} algorithm. Inference-based algorithms consider some or all pairs of data records, detecting dependencies that do not hold, and inferring those that do hold. Hybrid algorithms like HyMD combine these strategies.

Let us consider an example. In order to reduce the number of natural decision boundaries which we have to consider, we will modify our original Airline Flights dataset (Table~\ref{tbl:routes}). Firstly, we keep only the \enquote{From}, \enquote{To}, and \enquote{Distance} attributes. Secondly, we keep only the first four rows. The resulting dataset is presented in Table~\ref{tbl:routesCondensed}.

Furthermore, we will consider only column matches that map columns to themselves. For \enquote{From} and \enquote{To} attributes we will use normalized Levenstein distance as the similarity measure, and for the \enquote{Distance} attribute we will use normalized metric distance. It is calculated according to the following formula: $$d(x, y) = 1 - \frac{|x - y|}{maxDistance},$$ where $maxDistance$ is the greatest value for the \enquote{Distance} attribute in the dataset. For simplicity, all values are arithmetically rounded to one decimal place.

\begin{table}
\centering
\caption{Airline flights (dataset excerpt)}
\label{tbl:routesCondensed}
\begin{tabular}{l|l|l|c}
id & From             & To            & Distance (km)  \\ 
\hline \hline
1  & Saint-Petersburg & Helsinki      & 315            \\
2  & St-Petersburg    & Helsinki      & 301            \\
3  & Moscow           & St-Petersburg & 650            \\
4  & Moscow           & St-Petersburg & 638            \\
\end{tabular}
\end{table}

The similarity table for those column matches is given in table~\ref{tbl:similarities}.

\begin{table}
\centering
\caption{Airline flights similarity table}
\label{tbl:similarities}
\begin{tabular}{l|l|l|c}
Records & From & To  & Distance \\ 
\hline \hline
1, 2    & 0.9  & 1.0 & 1.0     \\
1, 3    & 0.1  & 0.2 & 0.5     \\
1, 4    & 0.1  & 0.2 & 0.5     \\
2, 3    & 0.1  & 0.2 & 0.5     \\
2, 4    & 0.1  & 0.2 & 0.5     \\
3, 4    & 1.0  & 1.0 & 1.0     \\
\end{tabular}
\end{table}

\begin{figure*}
    \centering
    \includegraphics[width=\linewidth]{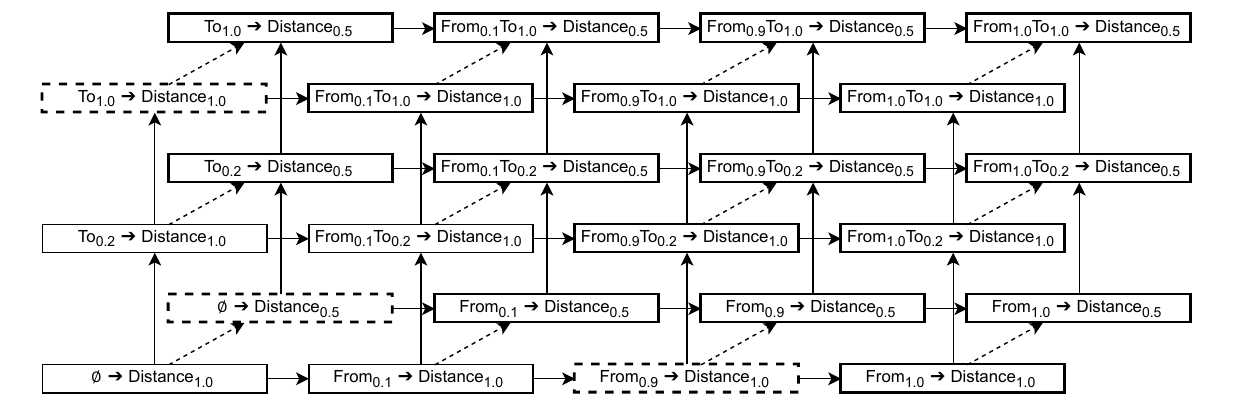}
    \caption{The search space lattice for matching dependencies of our example (Table~\ref{tbl:routesCondensed}) with \enquote{Distance} attribute in the RHS. Solid lines denote specializations of a single attribute in LHS, dashed lines~--- of the RHS. Bold rectangles denote holding MDs, dashed~--- minimal holding MDs.}
    \label{fig:lattice}
\end{figure*}

\textbf{Lattice traversal} incrementally builds search space consisting of MDs, starting from the most general MD and specializing it. Conceptually, there is a separate lattice for each attribute in the RHS of a MD, but implementations often process multiple lattices simultaneously. For our running example, consider the lattice for the \enquote{Distance} attribute given in Fig.~\ref{fig:lattice} (trivial and non-disjoint MDs are omitted). The bottom left MD (called \enquote{root}) is the most general and top right is the most specialized. Here, the root MD states that all values of \enquote{Distance} attributes must be equal no matter what the other attributes are. Top right MD states that, if \enquote{From} and \enquote{To} attributes are equal, \enquote{Distance} attribute can be anything (similarity $0.5$ is the lowest natural decision boundary in this dataset, it is impossible for the \enquote{Distance} attribute values to differ more).

We can note that the root MD is the least likely to be valid on a dataset (but if it holds, it is guaranteed to be minimal and no other minimal MDs are possible), and that the top right MD always holds. Each arrow in Fig.~\ref{fig:lattice} represents a single step of specialization, involving the next natural decision boundary (increasing for LHS, decreasing for RHS)~--- the MD at the start of the arrow subsumes the MD at the end. The main idea of lattice traversal is to inspect the lattice from the root (the least element) to the greatest element validating MDs. If a dependency is found to be invalid, it is removed from the lattice, and all its specializations that are not subsumed by other MDs in the lattice (i.e. minimal) are added. All MDs that are subsumed by a holding MD are guaranteed to hold, so the lattice can be pruned very efficiently. Intuitively, minimal holding MDs are located at the border between the bottom-left region of non-holding MDs and the top-right region of holding but non-minimal MDs. Minimal holding MD can only have incoming arrows from non-holding ones.

\textbf{Inference from record pairs} considers a set of MDs and iterates over the record pairs from the dataset, detecting all MDs that are violated by the current record pair. During the process, we maintain a set of candidate MDs $\Phi$, which is initialized with the root MD at the start. All MDs from $\Phi$ are checked against the current record pair. For those that are violated, we create several more MDs by either lowering the RHS decision boundary so that the new MD is not violated by the pair, or by raising an LHS decision boundary so that the pair is no longer matched by the new MD's LHS. We then add MDs that are minimal. When all record pairs are considered, $\Phi$ contains minimal holding MDs.

\subsection{Algorithm}

\begin{figure*}
    \centering
    \includegraphics[width=0.7\linewidth]{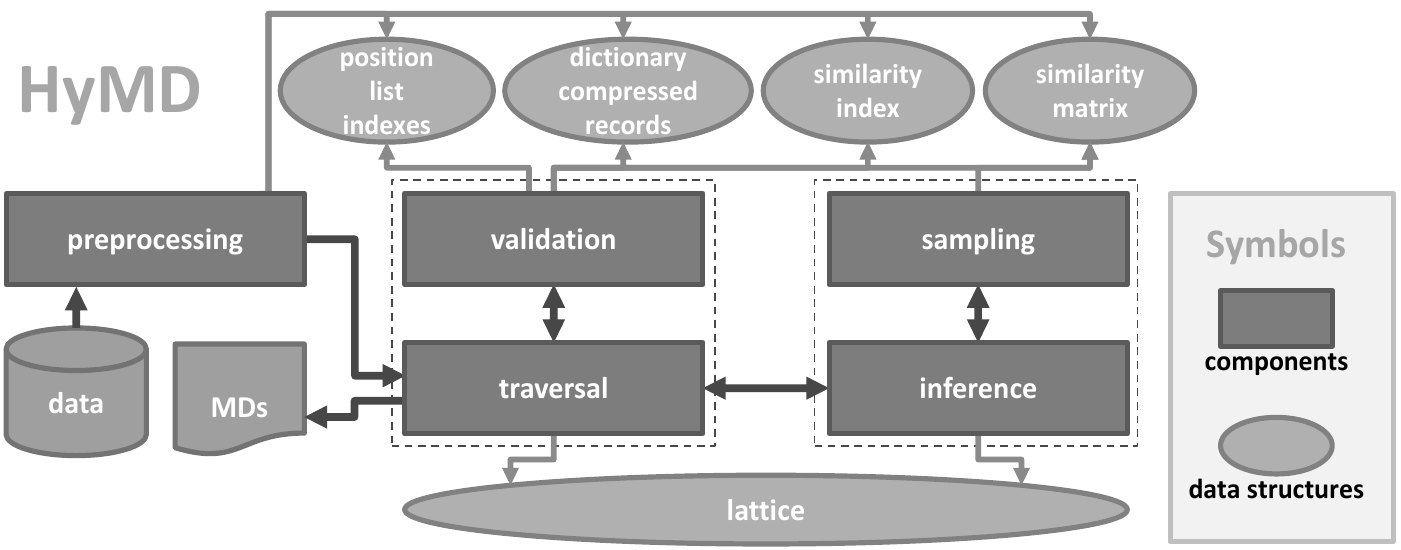}
    \caption{An overview of the HyMD algorithm, figure from~\cite{hymd}.}
    \label{fig:hymd}
\end{figure*}

An overview of the HyMD algorithm is depicted on Fig.~\ref{fig:hymd}. The algorithm uses several supporting data structures:

\begin{itemize}
    \item \textbf{Position list indexes (PLIs)}~--- for each column of both tables, a PLI keeps an array of sets of records (clusters) that share a value in that column.
    \item \textbf{Dictionary compressed records}~--- for both tables, this structure stores which value each record contains in every column. Values are represented by their numeric identifiers, which are indices of the relevant PLI clusters.
    \item \textbf{Similarity matrices}~--- for a column match, a similarity matrix stores information on how similar each value in the first column is to a value in the second column.
    \item \textbf{Similarity indexes}~--- for a column match, a similarity index maps a value from the first column and a natural decision boundary to the set of records that have a value in the second column that is at least as similar to the first value as the natural decision boundary indicates. Records are represented by their numeric identifiers, which are indices of the record in the Dictionary Compressed Records structure.
    \item \textbf{Lattice}~--- stores the set of minimal MDs that are assumed to hold in compact form, as well as providing some operations for the algorithm to use. It is implemented as a prefix tree, with the root node denoting an empty LHS and the path to each node specifying a unique LHS. Each node holds all RHS decision boundaries for every column match. Initially, the root MD is assumed to have the greatest possible RHS decision boundaries that are then lowered during algorithm operation.
\end{itemize}

The algorithm starts in the preprocessing stage by creating the first two structures.

The tables are read record-by-record and each value is added to a value-to-position mapping. The positions from the mapping are used as value IDs in Dictionary Compressed Records, while the record positions are stored in the corresponding clusters.

The algorithm then proceeds to calculate similarities and build similarity matrices and indexes. Note that HyMD has an option to exclude similarities that are below a user-specified threshold from consideration. In this case they are considered to be $0.0$ and are not stored in order to save space.

Having finished with preprocessing, the algorithm starts the discovery process.

First, the initial batch of record pairs is sampled, and the lattice is updated using inference from record pairs. Sampling in HyMD is quite straightforward: we pick a record from the first table and compare it to all records from the second table. We then use those comparison results to update the lattice. This step ensures lattice traversal starts with a lattice that is closer to its final state.

During lattice traversal, the algorithm picks up MDs from the lattice, then validates and specializes them until it stops, regardless of whether all MDs have been validated or not. If all MDs have been validated, the algorithm stops. Otherwise, interesting record pairs (recommendations) generated during this process are passed to the next phase, which uses them to update the lattice further.

The algorithm continues switching phases, always ending with lattice traversal, as only this phase can filter out insufficiently supported MDs.

In the end, the lattice contains all minimal interesting MDs that hold on the dataset.

\section{Proposed Improvements}\label{sec:improvements}

\subsection{Focused sampling}

The original sampling strategy proposed in~\cite{hymd} compares records one by one until inference from record pair becomes inefficient, according to a heuristic. The heuristics described in~\cite{hymd} and used in the original implementation are different. However, we have found our own sampling strategy, described below, to be superior to using either of them.

The strategy is in some way similar to the focused sampling proposed in~\cite{hyfd} but is adapted for MDs. We sample not by column, but by column match, choosing out of two available sub-strategies depending on its parameters until sampling for all column matches becomes inefficient. The efficiency of each column match is determined by the ratio of similarity sets revealing a violation to the total number of pairs compared. We start with an efficiency threshold of $0.01$, returning to the lattice traversal phase once the ratio falls below this threshold. The efficiency requirement is relaxed upon returning to the sampling stage.

\begin{figure}
    \centering
    \includegraphics[width=0.9\linewidth]{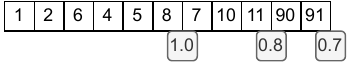}
    \caption{A possible representation of a similarity index.}
    \label{fig:slim-similarity-index}
\end{figure}

In place of comparing a PLI cluster's records to themselves in a sliding window manner, as proposed in~\cite{hyfd}, we make use of a structure created in the process of building similarity indexes. Note, for example, that a value that has a similarity of $0.8$ or greater to another value also has a similarity of $0.7$ or greater, meaning the record with that value will be present in both sets of the similarity indexes. It is thus possible to represent a similarity index as an array of record IDs with pointers to the start/end of the sets of records.

This structure is illustrated in Figure~\ref{fig:slim-similarity-index}. The array of record IDs is in the top row, and the marks for where record sets end are in the bottom row. The record set for similarity $1.0$ consists of records preceding the $1.0$ mark. The record set for similarity $0.8$ consists of records preceding the $0.8$, same for the record set for similarity $0.7$. Only the array of records is used during sampling.

\begin{figure}
    \centering
    \includegraphics[width=0.45\linewidth]{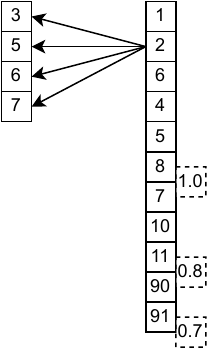}
    \caption{Full sampling illustration}
    \label{fig:full-sampling}
\end{figure}

We start with the first sub-strategy~--- full sampling. For each value in the first column, we compare all the records in its PLI cluster with a chosen record from the second column. We pick it using an index in the array of records from the similarity index, starting with 0, and then incrementing it for the next round of sampling. This way, record pairs with higher similarities for this column match are prioritized.

We illustrate the approach in Figure~\ref{fig:full-sampling}. During the second round of sampling for a column match, a record with ID 2 is compared with all records in the first table's cluster.

It is possible to go further and order the records in such a way that the similarities in other column matches during earlier rounds of sampling are higher, in a similar manner to HyFD's sorting every cluster so that records with equal values end up closer together. The way to achieve this is by sorting with a specific key: for each record we create a list of similarities, sorted lexicographically, using similar lists for other column matches as tie-breakers. We obtain this list of similarities by comparing a record from the second column with the relevant cluster's records. The exact column matches are not important, the only condition would be that, for all column matches, no two tie-breakers share places in the tie-breaker orderings. We have not found a significant benefit to doing that in our experiments, while it reduced performance significantly on larger datasets. Thus, we have opted not to sort the records further.

\begin{figure}
    \centering
    \includegraphics[width=0.9\linewidth]{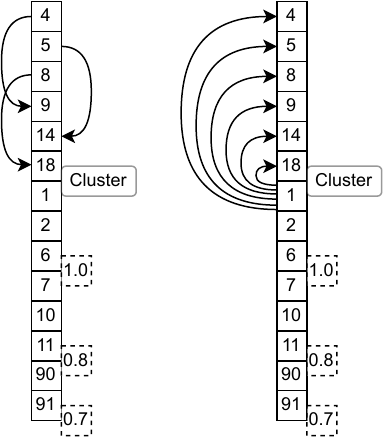}
    \caption{Short sampling illustration}
    \label{fig:short-sampling}
\end{figure}

Take note that, in the common case where there is only one table and every column is matched to itself, with all measures being symmetrical and outputting $1.0$ if values are equal ($\bigstar$), this would result in a lot of repeated similarity sets for the records contained in a cluster, quickly making record pair inference inefficient. In this case, a different sampling strategy is used. It first compares record pairs in a cluster, just like cluster windowing does in HyFD. After cluster records are exhausted, the regular full sampling is used.

Note that the last two conditions in $\bigstar$ are true for similarity measures used in practice, with the former one being implied by the definition in \cite{fan2008dependencies}. However, we make it explicit that we rely on those properties in our implementation.

This method is illustrated in Figure~\ref{fig:short-sampling}. In the first case (on the left) the parameter of sampling is 3, and the size of the cluster is 6, so the sampling process does a sliding window round with the window size of 3. In the second case, the parameter of sampling is 6, which is not less than the cluster's size, thus full sampling is used~--- record 1 is compared with all records in the cluster.

Getting the best ordering of the records in the cluster in the same sense as before can be reformulated as finding a path in a complete graph with certain properties, and the algorithm to find such a path can also be used to find a Hamiltonian path in any graph with polynomial-time transformation of the input, thus the problem is NP-hard. We still decided to sort the cluster in this case, but chose a relatively inexpensive sorting technique, i.e. the one used in HyFD.

With the sampling strategy described above, we have found that, if the algorithm switched phases after a single round of lattice traversal, it would spend a lot of time in the inference phase doing useless work as the lattice would be in the final state. We adapted the heuristic from HyFD: we only stop lattice traversal if the portion of invalidated MDs among all MDs is above a certain low value. After implementing focused sampling and this new heuristic, we have found significant reductions in running time.

\subsection{Ignoring irrelevant nodes when checking for generalizations}

Before reading further, take note of the fact that all RHSs' decision boundaries are stored in the lattice's node as an array. Setting a decision boundary in that array to $0.0$ is exactly the same as removing it from the lattice.

During both inference from record pairs and lattice traversal, after an MD has been invalidated, its specializations are added into the lattice, provided they are minimal. The only other time an MD is added to the lattice is during lattice traversal, when an MD that used to be there is invalidated. Still, we ensure that the newly added MD is minimal among all the ones present in the lattice by checking that the decision boundary is higher than the greatest decision boundary among all MDs in the lattice the LHSs of which generalize the LHS of the newly added MD.

When there is only one MD in the lattice, like at the start of either phase, it is obviously minimal among all MDs in the lattice. Thus, by a simple induction argument, we can prove that at any point during execution of all algorithms specified, be it lattice traversal, inference from record pairs, or the hybrid approach, all MDs stored in the lattice structure are minimal.

We can use this fact to our advantage when checking if the newly added MD is a generalization of another one. It should be noted that when a new MD is added using \texttt{addIfMin} (as described in section 5.2 of \cite{hymd}), its LHS is always a specialization of another MD that was in it beforehand, and that the RHS is the same. That previous MD used to be minimal among all others. During the addition process, the lattice holds exactly the same MDs as it used to when the old MD was there, except the decision boundaries of the invalidated MDs are lower. That means that the lattice MDs with LHSs that generalize the previously present MD are certainly not generalizations of the MD we are trying to add. Thus, these MDs do not need to be checked.

We have not found this optimization in other algorithms, but it is applicable in FD and UCC mining as well.

\subsection{More optimal LHS representation}

Following the example of HyFD, the original implementation of HyMD used an array of similarities to represent the LHS of an MD. This would result in a lot of checking for equality to $0.0$ during procedures that traversed the lattice tree, as $0.0$ decision boundaries are left out of the tree during the algorithm's operation.

In the new implementation, a different structure to represent LHSs is used. Instead of an array of similarities, we use an array of pairs, where the first element is the distance from the previous column match, and the second one is a non-zero similarity in that LHS. With this, finding the next non-zero similarity amounts to incrementing a pointer.

Despite the simplicity of the change, it led to unexpectedly significant running time improvements.

\subsection{Using decision boundary indices}

In the initial implementation, natural decision boundaries are stored directly in the similarity matrices, similarity indexes, and in the list of LHS decision bounds that are to be checked. However, we have found that representing decision boundaries as their indices in the list of all natural decision boundaries of a column match allows for additional optimizations. This comes at a comparably small cost of having to determine the index of each decision boundary and additional complexity in the implementation.

Firstly, \cite{hymd} mentioned an optimization to the \texttt{getLowerBoundaries} method, where it would stop traversing the lattice in search of higher decision boundaries if all decision boundaries become $1.0$. However, this situation is not possible if the lattice is only used during algorithm execution.

That method is only ever used as part of the validation process, and only MDs that are actually present in the lattice are validated. Recall from a prior section that the lattice structure has an invariant where only minimal MDs of those assumed to be holding at some point during execution are in the lattice. If we find that any one RHS boundary reached $1.0$, that means there was an MD stored in the lattice with a more general LHS and a higher RHS decision boundary than the one being validated, which violates the stated invariant.

Using this same invariant, we may state a different condition for this optimization. We abort the operation once all indices reach exactly one less than the RHS decision boundary indices being validated, which is, again, in the lattice. Although this can be applied even if we are working with decision boundaries directly by stopping once the decision boundary previous to the ones being validated are reached, working with indices makes implementation trivial.

We have observed this condition work during execution. On the \texttt{adult} dataset, this search for the highest decision boundaries was skipped entirely 90 times out of 106 validations. On the \texttt{flight} dataset, there were 14540 validations in total, with 7748 not having to execute the procedure at all, and 422 stopping early.

\subsection{Other technical improvements}

There are a number of technical improvements that we believe are interesting, but not enough for them to have a separate section.

Firstly, for each RHS array in every node, we keep count of non-zero elements in the node. The idea here is to avoid loading the part of memory where actual values are stored if we know there are no useful (non-zero) elements in the array. This reduces cache usage.

Another notable change is that we sort column matches by the number of LHS decision boundaries searched, ascending. This leads to reduced memory usage for the lattice tree, as fewer pointers to next nodes need to be stored in the deeper nodes, making a larger portion of the lattice fit into cache.

We also remove nodes from the lattice once they become empty: all values are zero in the RHS array and there are no child nodes. It also reduces memory usage, but, more importantly, some methods, like \texttt{FindViolated} and generalization checking methods, don't visit those nodes, reducing running time.

There was another optimization idea, which is to have a special procedure for adding several RHSs with the same LHS at once. The rationale was that loading all the nodes that need to be visited into cache used up a lot of space, but the nodes were always the same. This did not improve running time, as adding several RHSs at once is exceedingly rare with the tested use cases. That might not hold true in the future algorithm mentioned before if several measures are used for one RHS column pair.

\section{Evaluation}\label{sec:eval}

To evaluate our techniques, we have developed our HyMD implementation inside Desbordante and experimentally compared it with the existing HyMD implementation (https://github.com/HPI-Information-Systems/metanome-algorithms/tree/master/HyMD) written in Java.

\subsection{Methodology}

We posed the following research questions:

\begin{itemize}
    \item[RQ1] Is it possible to outperform the existing implementation just by reimplementing MD discovery algorithm in C++?
    \item[RQ2] What is the performance impact of our proposed optimizations?
    \item[RQ3] What are the memory savings of the C++ implementation?
\end{itemize}

To answer these RQs, we have conducted a number of experiments:

\begin{enumerate}
    \item The overall evaluation of three implementations: vanilla HyMD (Metanome), vanilla HyMD (Desbordante), optimized HyMD (Desbordante).
    \item A study of the optimized HyMD implementation (Desbordante), in which we explored its scalability over rows and columns.
    \item An examination of memory consumption for all three implementations.
\end{enumerate}

Due to results being relatively stable, each experiment was repeated 5 times, the average of the results was reported.

Experiments were performed using a number of datasets, which are presented in Table~\ref{tab:datasets} alongside their characteristics.

Experiments were performed using the following hardware and software configuration. Hardware: Intel(R) Core(TM) i7-9750H CPU @ 2.60GHz (6 cores, 12 threads), 16 GiB RAM. Software: Arch linux, glibc 2.40, gcc 14.2.1, OpenJDK Runtime Environment (build 1.8.0\_422-b05).

All runs were performed on one relation, with every column matched to itself using a single similarity measure: Monge-Elkan for CORA and Levenshtein for the others.

\begin{table*}[!t]
    \centering
    \caption{Datasets}
    \label{tab:datasets}    
    \begin{tabular}{|c|c|c|c|c|c|}
        \hline
        Dataset & Columns & Rows & Size & LHSs & \#MDs \\
        \hline \hline
        \rowcolor{LightGray}
        restaurants & 6 & 864 & 63.5 KB & $10^{6}$ & 7 \\
        \hline
        adult & 15 & 32561 & 3.6 MB & $10^{7}$ & 111 \\
        \hline
        \rowcolor{LightGray}
        CIPublicHighway50k & 18 & 50000 & 3.7 MB & $10^{7}$ & 280 \\
        \hline
        flights & 38 & 1000 & 190.7 KB & $10^{15}$ & 36637 \\
        \hline
        \rowcolor{LightGray}
        breast\_cancer & 30 & 569 & 120.4 KB & $10^{19}$ & 29079 \\
        \hline
        notebook\_ascii & 14 & 337 & 194.3 KB & $10^{19}$ & 32094 \\
        \hline
        \rowcolor{LightGray}
        CORA & 16 & 1879 & 400.3 KB & $10^{37}$ & 24155 \\
        \hline
    \end{tabular}
\end{table*}

\begin{table*}[!t]
    \centering
    \caption{Total running times (ms)}
    \label{tab:total-runtimes}    
    \begin{tabular}{|c|c|c|c|c|c|c|}
        \hline
        Dataset & Metanome (M) & Desbordante (R) & Desbordante optimized (O) & M/R & R/O & M/O \\
        \hline \hline
        \rowcolor{LightGray}
        adult & 33844 & 7854 & 4465 & 4.31 & 1.76 & 7.58 \\
        \hline
        breast\_cancer & 1135762 & 120612 & 16653 & 9.42 & 7.24 & 68.20 \\
        \hline
        \rowcolor{LightGray}
        CIPublicHighway50k & 247719 & 107900 & 49734 & 2.30 & 2.17 & 4.98 \\
        \hline
        CORA & 4772748 & 4588743 & 28011 & 1.04 & 163.82 & 170.39 \\
        \hline
        \rowcolor{LightGray}
        flights & 202311 & 43010 & 2914 & 4.70 & 14.76 & 69.43 \\
        \hline
        notebook\_ascii & 129708 & 11880 & 4085 & 10.92 & 2.90 & 31.75 \\
        \hline
        \rowcolor{LightGray}
        restaurants & 2510 & 163 & 38 & 15.40 & 4.29 & 66.05 \\
        \hline
    \end{tabular}
\end{table*}

\begin{table*}[!t]
    \centering
    \caption{Detailed running times (ms)}
    \label{tab:detailed-runtimes}    
    \begin{tabular}{|c|c|c|c|c|c|c|c|c|c|}
        \hline
        Dataset & \multicolumn{3}{|c|}{Preprocessing} & \multicolumn{3}{|c|}{Execution} & \multicolumn{3}{|c|}{Total} \\
        \hline
         & M & R & O & M & R & O & M & R & O \\
        \hline \hline
        \rowcolor{LightGray}
        adult & 20992 & 5957 & 3217 & 11551 & 1879 & 1231 & 33844 & 7854 & 4465 \\
        \hline
        breast\_cancer & 2331 & 160 & 98 & 1133191 & 120452 & 16554 & 1135762 & 120612 & 16653 \\
        \hline
        \rowcolor{LightGray}
        CIPublicHighway50k & 215248 & 94526 & 37596 & 30323 & 13374 & 12018 & 247719 & 107900 & 49734 \\
        \hline
        CORA & 8271 & 1814 & 1222 & 4764118 & 4586687 & 26789 & 4772748 & 4588743 & 28011 \\
        \hline
        \rowcolor{LightGray}
        flights & 1835 & 27 & 9 & 200427 & 42982 & 2905 & 202311 & 43010 & 2914 \\
        \hline
        notebook\_ascii & 3002 & 923 & 149 & 126641 & 10929 & 3935 & 129708 & 11880 & 4085 \\
        \hline
        \rowcolor{LightGray}
        restaurants & 2247 & 151 & 34 & 249 & 8 & 4 & 2510 & 163 & 38 \\
        \hline
    \end{tabular}
\end{table*}

\begin{figure}
    \centering
    \includegraphics[width=0.9\linewidth]{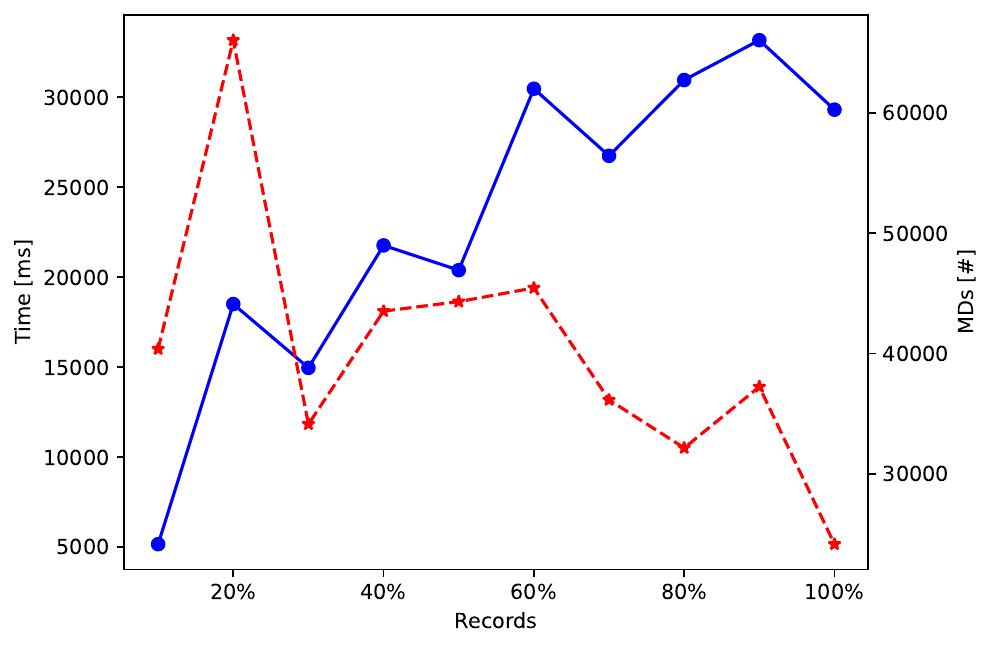}
    \caption{Scaling the number of records}
    \label{fig:record-scaling}
\end{figure}

\begin{figure}
    \centering
    \includegraphics[width=\linewidth]{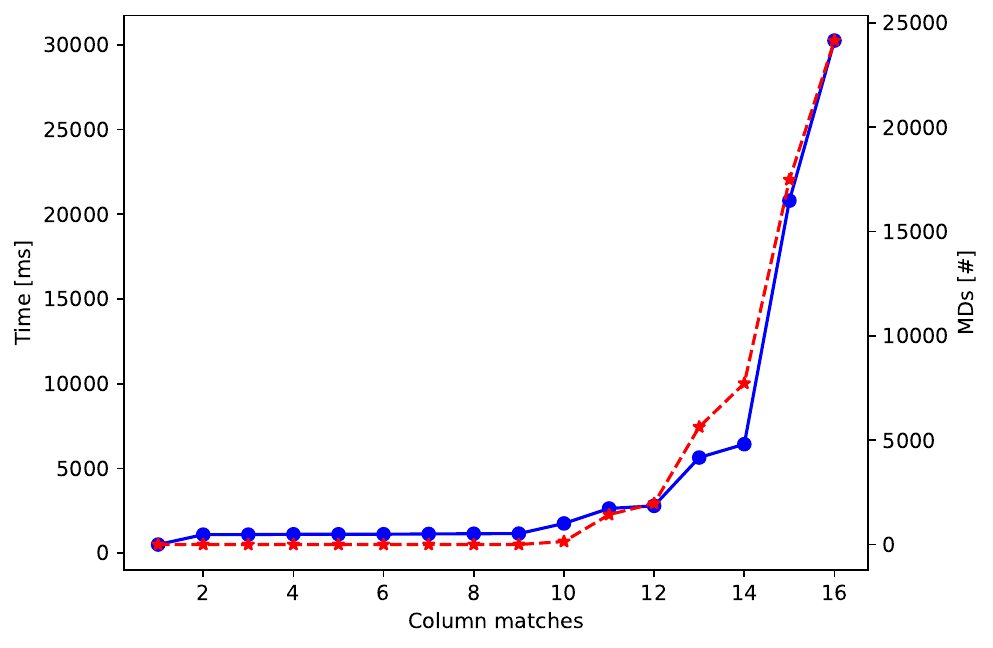}
    \caption{Scaling the number of column matches}
    \label{fig:column-match-scaling}
\end{figure}

\begin{table*}[!t]
    \centering
    \caption{Memory (KB)}
    \label{tab:memory}    
    \begin{tabular}{|c|c|c|c|c|c|c|c|}
        \hline
        Dataset & Metanome (M) & Desbordante (R) & Desbordante optimized (O) & M/O \\
        \hline \hline
        \rowcolor{LightGray}
        adult & 1856592 & 153120 & 159752 & 11.62 \\
        \hline
        breast\_cancer & 2313360 & 483412 & 145852 & 15.86 \\
        \hline
        \rowcolor{LightGray}
        CIPublicHighway50k & 2240160 & 1015840 & 1208060 & 1.85 \\
        \hline
        CORA & 3022900 & 4080900 & 493228 & 6.13 \\
        \hline
        \rowcolor{LightGray}
        flights & 1887912 & 666784 & 124884 & 15.12 \\
        \hline
        notebook\_ascii & 1737012 & 189656 & 66212 & 26.23 \\
        \hline
        \rowcolor{LightGray}
        restaurants & 663368 & 8320 & 15200 & 43.64 \\
        \hline
    \end{tabular}
\end{table*}

\subsection{Experiments}

\textbf{Experiment 1.} In this experiment, we studied the overall performance of three implementations: vanilla HyMD (Metanome), vanilla HyMD (Desbordante), and optimized HyMD (Desbordante). The results are presented in Table~\ref{tab:total-runtimes}. In most cases, the optimized implementation was an order of magnitude faster than the Metanome implementation.

Furthermore, we have separately measured the running times of each individual phase. These results are shown in Table~\ref{tab:detailed-runtimes}.

Note that unlike~\cite{hymd}, we did not limit the number of LHS decision boundaries for the CORA dataset, which meant the number of total possible LHSs for it was on the order of undecillions, even though the number of records was only 1879. For this dataset, inference from record pairs is obviously the more fitting algorithm. Both the original implementation and the rewrite performed poorly in this case, as due to the switching heuristic used, inference from record pairs stopped too early. The optimized version used a different sampling mechanism and switched much later, along with inspecting more promising record pairs, resulting in a massive 170x speedup. Note that reimplementing barely gave any speedup. For this dataset, the vast majority of execution time is spent in the implementation equivalent of the \texttt{FindViolated} method as defined in \cite{hymd}. We presume that, in this case, the JIT compilation worked particularly well, which could explain the small difference.

It should be noted that the preprocessing time decreased for the optimized version on all datasets relative to the vanilla version. The optimizations described before concern only the execution phase of the algorithm, so these numbers only decreased due to rewriting and using faster metric functions. There are many other ways to exploit the particular circumstances of similarity measure calculation, but they were mostly not implemented in the optimized version. This area has a lot of potential for future optimization.

If we inspect the detailed table for the CIPublicHighway dataset, we will see that the decrease in total runtime was due to the decrease in preprocessing time, with the execution time staying almost the same, with just an 11\% decrease. On this dataset, the new sampling helped only a little. However, for all other datasets, the execution time was reduced dramatically, which proves this sampling method viable.

If we inspect execution time, we will notice that, aside from the CORA dataset, simply reimplementing the algorithm in C++ decreased it.

\textbf{Experiment 2.} In the second experiment, we studied the scalability of the optimized algorithm over rows and columns. The CORA dataset was used for this experiment.

The first part was performed on different versions of CORA that contain a fixed percentage of randomly selected rows. The results are shown in the Figure~\ref{fig:record-scaling}. The dashed red line represents the number of discovered MDs, while the solid blue one plots the running time. As can be seen, the execution time does not always increase with the number of records. This occurs because the increase in the number of rows affects the algorithm in both ways: while it slows down the validation procedure, it can also speed up the inference phase by providing useful MD violations. However, in general, the time increases and the number of found MDs decreases with growing number of records. Note that this plot is different from the Figure 5 of the original HyMD paper, which contained a similar experiment. This could happen because the authors of~\cite{hymd} added records to those selected for the previous step for all steps, while we sampled them randomly.

The second part of this experiment addressed the scalability of the algorithm over the number of columns. In this case, the size of the search space grows exponentially, and so does running time. The results are presented in Figure~\ref{fig:column-match-scaling}, the same way as in the first part of the experiment. It is clear that, as expected, the time grows exponentially, and so does the number of discovered MDs.

\textbf{Experiment 3.} Here we studied how the memory consumption differed between the three versions of the algorithm. The results are presented in Table~\ref{tab:memory}.

First, we compared the vanilla versions. This experiment demonstrated a reduction in memory consumption on all but one (CORA) datasets. It ranged from 2.2x to 80x, with 16x being the average.

Then, we compared the optimized version of Desbordante with Metanome. In all cases, the Desbordante implementation consumed significantly less memory, more than 10 times less in most cases. The outlier was the CIPublicHighway50k dataset, which exhibited only 1.85x memory consumption reduction. Overall, the reduction ranged from 1.85x to 43.64x with 17x being the average.

\textbf{Discussion.} Our experiments consistently show that the answer to RQ1 \enquote{Is it possible to outperform existing implementation just by reimplementing MD discovery algorithm in C++?} is positive, despite the actual speedup being highly dependent on dataset. This speedup seems to be unrelated to JVM options, since in our previous work~\cite{desbordante} we had extensively experimented with a large number of options and found the default settings to be the best.

Experiment 1 shows that proposed optimizations can provide additional speedup of up to 163x, again, depending on a dataset, but consistently significant, which answers the RQ2 \enquote{What is the performance impact of our proposed optimizations?}. The highest speedup can be achieved on datasets with many values that are similar but not equal (such as CORA), but such datasets are inherently hard for MD discovery due to the high amount of possible natural boundaries and LHSs.

RQ3 \enquote{What are the memory savings of C++ implementation?} is more interesting, since, despite the C++ implementation consistently showing significantly less memory usage than Java implementation in Metanome, there is no direct correlation between memory consumption of vanilla and optimized versions. This can be explained by the fact that there are many factors influencing the memory usage in the algorithm and they non-trivially depend on the optimizations used. The most noticeable one on the datasets with many possible MDs is the size of the lattice, in which case the improved heuristic for sampling can significantly reduce lattice memory footprint. However, on those with fewer holding MDs, the similarity indexes can take up a lot more space than the lattice. In the optimized version, an array with the same number of elements as the largest set for a value from the similarity index is stored for the needs of the sampling procedure, which may explain the higher memory usage in some cases.

\section{Future work}\label{sec:future}

We have noticed that, for all datasets, more than half of the execution time was spent waiting for memory in lattice-related methods. It is thus clear that improving the memory layout of the lattice tree structure is of utmost importance for reducing running time.

Another way to improve running time is by utilizing the available throughput. Currently, all lattice-related operations are designed to be performed in a single thread. Making use of available concurrency seems to be another avenue of improvement.

Aside from memory access times, another potential issue seems to be memory consumption. Similarity indexes are, by their nature, rather large structures, so some thought should be put into reducing their size. For example, we were unable to process the NCVoters and Amazon-Walmart datasets with 32 GB of available memory because the similarity indexes took up too much space. Compressing similarity indexes in various ways, such as using value IDs instead of record IDs, looks to be an important consideration.

In addition to potential improvements, there is a possibility of using the algorithm with some changes to discover a wider class of dependencies. The theory of MDs can be reformulated to allow not only numbers in the $[0.0, 1.0]$ range as results of similarity measures, but any type with a total order. With these changes, the class of dependencies the algorithm can discover will directly subsume metric functional dependencies without a need for further transformations.

For clarity, it is still possible to extract those dependencies from the dataset using HyMD by choosing particular measures, then filtering and transforming the results, although it will be inefficient. A more efficient mining of this type of dependency will require a modified algorithm.

\section{Conclusion}\label{sec:concl}

In this paper, we have presented a number of optimizations for HyMD~--- the state-of-the-art algorithm for discovery of matching dependencies. Evaluation demonstrated that our techniques can provide up to a 170x speedup over the existing implementation. We have also reduced memory consumption by more than tenfold on average.

The resulting algorithm implementation has become a part of Desbordante~--- a science-intensive, high-performance, and open-source data profiler (\url{https://github.com/Desbordante/Desbordante-core/}, PR 456). Thus, efficient discovery of matching dependencies has become available to the broader public through the Python interface.

Matching dependencies are arguably one of the most important patterns. Their efficient discovery is of utmost importance, as it will allow users to solve various data quality problems. There are several possible applications:

\begin{itemize}
    \item Data integrity maintenance within DBMSes. The expectations in regards to flexibility of data consistency rules are constantly rising, leading to demand for novel ways of specifying various constraints. Matching Dependencies can be used as a kind of approximate primary key useful for records containing complex (e.g. address strings) or dirty data, or, alternatively, as a way to define a relative constraint over a table or a pair of tables.
    
    \item Master Data Management applications~\cite{10.5555/2815504, 10.5555/1457711}, typically implemented as stand-alone applications. They address various data quality problems such as detecting inexact duplicates, performing record or schema matching. In this case, Matching Dependency discovery can be directly used for designing programs or scripts which will be used for the aforementioned tasks.
    
    \item Data exploration applications (data profilers, such as Desbordante) which help users understand their data as well as provide them with insights and hypotheses concerning data domain. Matching Dependencies are very expressive and provide one-of-a-kind type of information.
\end{itemize}

Overall, we hope that supporting MDs in Desbordante will bring MDs closer to practical use.

\section*{Acknowledgments}

We would like to thank Vladislav Makeev for his help with the preparation of the paper.

\balance

\bibliographystyle{IEEEtran}

\bibliography{FRUCTexample}

\end{document}